\documentclass[aps,twocolumn,superscriptaddress]{revtex4-2}
\usepackage{amsmath,amssymb}
\usepackage{graphicx}
\usepackage{comment}
\usepackage{color}

\begin{document}

\title{Spatiotemporal Structures of Parametrically Driven Nonlinear Lattices}

\date{\today}
\author{Yu Funami}
\affiliation{Department of Earth and Space Science, Graduate School of Science, The University of Osaka, Osaka 560-0043, Japan}
\author{Kazushi Aoyama}
\affiliation{Department of Earth and Space Science, Graduate School of Science, The  University of Osaka, Osaka 560-0043, Japan}
\affiliation{Graduate School of Advanced Science and Engineering, Hiroshima University, Higashihiroshima 739-8521, Japan}

\begin{abstract}
We theoretically investigate the effects of parametric driving on the one-dimensional Frenkel–Kontorova model, a nonlinear many-body lattice system.
It is numerically found that a parametric vibration induces spatiotemporal ordering characterized by the subharmonic frequency $\omega_{\mathrm{ex}}/2$ and wavenumber $k$ which takes nontrivial values depending on the vibration strength $P_{\mathrm{ac}}$ and frequency $\omega_{\mathrm{ex}}$. With increasing $P_{\mathrm{ac}}$, $k$ gradually increases and discontinuously jumps up to $k=\pi$. 
Based on an extended linear stability analysis, we show that the former $k$ resonance (the latter $\pi$ resonance) corresponds to the unstable (stable) mode and that the $k$ resonance is made possible by the interplay between the parametric amplification and the collective fluctuation developing through the nonlinear effect.
\end{abstract}

\maketitle

Parametric driving is one of the most fundamental concepts in nonlinear and nonequilibrium collective dynamics, and it appears in a wide range of physical systems such as vibrated liquids hosting Faraday waves \cite{Faraday1831, Benjamin1954, Porter2004, Dinesh2023}, nonlinear optical systems \cite{Staliunas2013,Ripoll1999,Kivshar1998}, Bose–Einstein condensates \cite{Fu2018,Nguyen2019,Nicolin2007,Engels2007}, and micro- and nanoelectromechanical resonators \cite{Buks2002,Lifshitz2003,Lifshitz2008}. 
The parametric driving is characterized by its strength $P_{\rm ac}$ and frequency $\omega_{\rm ex}$.
When the strength $P_{\rm ac}$ exceeds a threshold value, the systems start to exhibit parametric resonance where oscillations of system degrees of freedom are amplified with their frequency fixed to be $\omega_{\rm ex}/2$ \cite{Kovacic2018,Parametric2023}. 
As the oscillation amplitude increases, nonlinear effects become more dominant, eventually leading to collective resonant phenomena such as synchronization and pattern formation \cite{Cross1993,Reyes2024,Kivshar19920901,Citro2015,Nicolaou202107,Chen2007,Denardo1995,Huang1993,Kivshar19921001,Denardo1992,Huang1994,Chen1994,Denardo1990,Bena2002,Newman1999}.
Although the parametric resonance has extensively been studied so far, full understanding of the collective dynamics is still a challenging issue.

\begin{figure}[b]
\includegraphics[width=0.8\columnwidth]{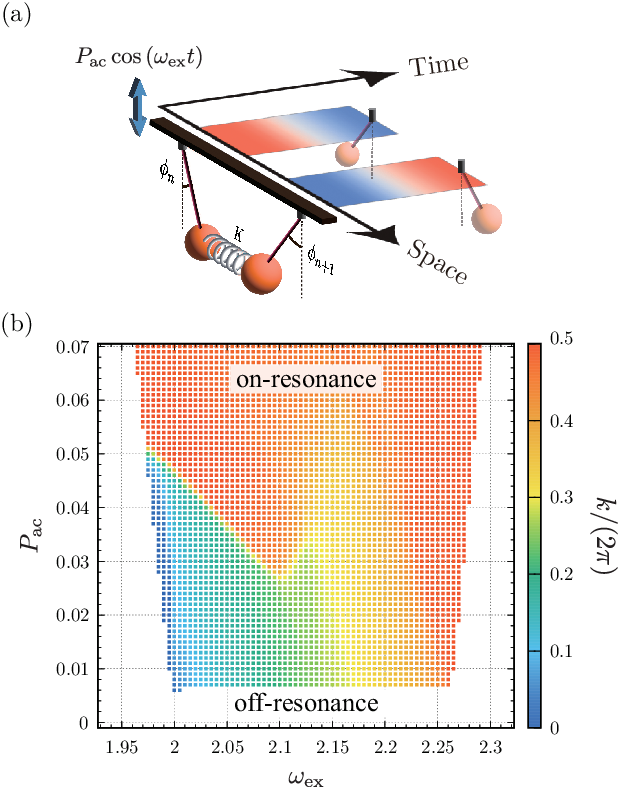}
  \caption{
  (a) Schematic of the $\pi$ resonance appearing in the one-dimensional FK model with the parametric driving, coupled oscillators under vertical vibration with strength $P_{\mathrm{ac}}$ and frequency $\omega_{\mathrm{ex}}$, where $\phi_n$ is the phase of the $n$th oscillator, and red (blue) regions in the space-time plane indicates positive (negative) values of $\phi_n$.
  (b) $P_{\mathrm{ac}}$ and $\omega_{\mathrm{ex}}$ dependence of a spatiotemporal (ST) structure with wavenumber $k$. The color represents $k/2\pi$ in the resonant state, while the white region denotes the off-resonant state. The result is obtained for the rest initial condition.
  }
  \label{fig:Phase_diagram}
\end{figure}

Recently, periodically driven many-body systems have attracted renewed interest in the context of a classical discrete time crystal \cite{Else2020, Zaletel2023, Heugel2023,Machado2023, Sarkar2026,Heugel2019,Yao2020,Nicolaou202105,Thomas2024}.
Since in the parametric resonance, individual system frequencies are collectively synchronized to be $\omega_{\rm ex}/2$, such a subharmonic resonance with many-body correlation can be interpreted as spontaneous breaking of discrete time-translation symmetry. 
One example of such an exotic state is a $\pi$-resonance in a parametrically driven Frenkel–Kontorova (FK) model, a nonlinear many-body oscillator system \cite{Denardo1992,Denardo1995,Chen2007,Yao2020,Huang1993}.
The $\pi$-resonance is schematically illustrated in Fig. \ref{fig:Phase_diagram}(a); in elastically coupled oscillators under a vertical parametric vibration, the subharmonic resonant state is accompanied by an antiphase spatial pattern with the wavenumber $\pi$ in which the phases of neighboring oscillators take opposite signs. 
This spatiotemporal (ST) structure is characteristic of the discrete lattice system and is stable in the wide range of the parameter space spanned by $\omega_{\rm ex}$ and $P_{\rm ac}$ \cite{Denardo1995}.
Although another type of the spatial pattern is reported to appear just near the onset of parametric instability \cite{Chen2007}, systematic search for ST structures has not been done so far.

In this work, we theoretically investigate the ST structure formation in the parametrically driven FK model, putting an emphasis on the possibility of other ST structures and the mechanism if they actually appear.
We will numerically demonstrate that for moderate driving strength $P_{\rm ac}$, nontrivial ST resonant sates with wavenumber $k\neq 0$ emerge [see the colored region in Fig. \ref{fig:Phase_diagram}(b)] and that with increasing $P_{\rm ac}$, $k$ gradually increases and discontinuously jumps up to $k=\pi$. By performing an extended linear stability analysis, we show that the $k$ and $\pi$ resonances are caused by different mechanisms.

\begin{figure}[t]
  \centering
  \includegraphics[width=\columnwidth]{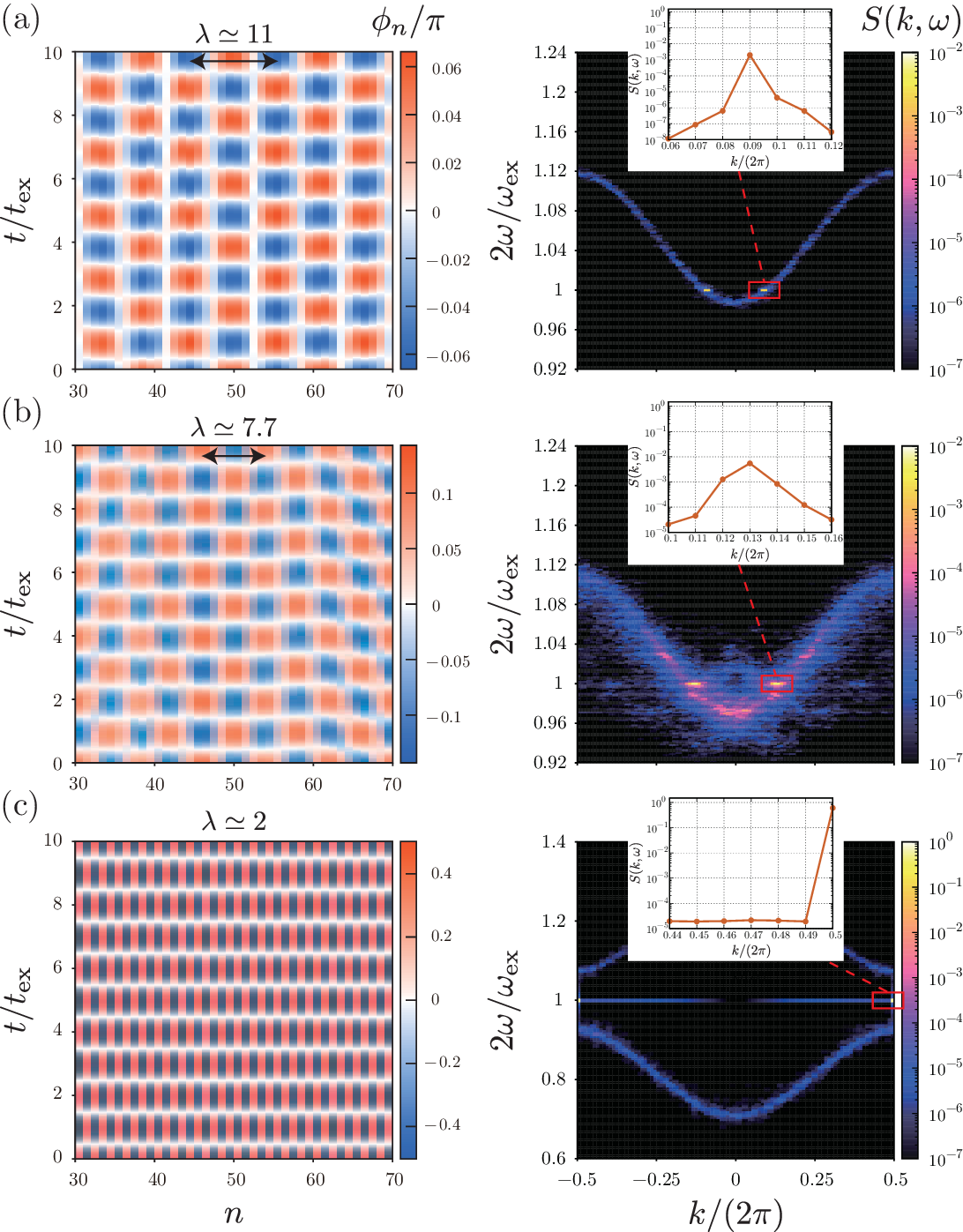}
  \caption{
  ST structures (left panels) and the corresponding Fourier spectra (right panels)  for $\omega_{\mathrm{ex}}=2.02$ :  (a) $P_{\mathrm{ac}}=0.007$, (b) $P_{\mathrm{ac}}=$ 0.024 , and (c) $P_{\mathrm{ac}}=0.05$. In the left panels, horizontal and vertical axes denote the site index $n$ and time $t / t_{\text {ex }}$, respectively, and the color represents the value of $\phi_n / \pi$. In the right panels, the color represents the spectrum intensity $S(k,\omega)$, and the insets show zoomed views near the Fourier peaks at $2\omega/\omega_{\rm ex}=1$.
  }
  \label{fig:ST}
\end{figure}
We start from the one-dimensional FK model. As shown in Fig. \ref{fig:Phase_diagram}(a), it describes elastically-coupled classical oscillators; a phase degree of freedom $\phi_n$ on the $n$th lattice site is interacting its nearest neighbor phases $\phi_{n-1}$ and $\phi_{n+1}$ with coupling constant $K$, being subject to an on-site potential in the form of $P\cos(\phi_n)$ \cite{Braun1998}. In the presence of the parametric drive corresponding to the vertical vibration in Fig. \ref{fig:Phase_diagram}(a), the on-site potential $P$ is time dependent, and then, the equation of motion is given by
\begin{equation}\label{Eq:FK_EOM}
\begin{aligned}
	\ddot{\phi}_n=&-\gamma\dot{\phi}_n-P(t)\sin \phi_n\\
	&+K(\phi_{n+1}-2\phi_n+\phi_{n-1}) +\xi_n
\end{aligned}
\end{equation}
with
\begin{equation}
\label{Eq:Parametric_Driving}
P(t)=\omega_0^2+P_{\mathrm{ac}}\cos\left(\omega_{\mathrm{ex}}t\right),
\end{equation}
where $\dot{\phi}_n=\frac{d}{dt}\phi_n$, $\gamma$ is a damping coefficient, and $\xi_n(t)$ is a Gaussian noise satisfying $\left\langle\xi_n(t) \xi_m\left(t^{\prime}\right)\right\rangle=2 \gamma T \delta_{n m} \delta\left(t-t^{\prime}\right)$ with noise strength $T$  \cite{Yao2020}. The mass of each oscillator and the lattice spacing are both set to be unity. In Eq. (\ref{Eq:Parametric_Driving}), the time-dependent term describes the effect of the external vibration with its strength $P_{\rm ac}$ and frequency $\omega_{\rm ex}$, whereas $\omega_0$ describes the static on-site potential.

In the case of a single oscillator without noise, i.e., $K=0$ and $T=0$, it is well established that the parametric resonance occurs for $\omega_{\mathrm{ex}} \sim 2\omega_0$ \cite{Kovacic2018, Parametric2023}.
In the many-body case of $K\neq0$, it is naively expected that all the oscillators are synchronized to exhibit the same parametric oscillation as that in the single-body case, but this is not the case. Due to the many-body interaction, the resonant condition is modified \cite{Ameye2025}, which results in the emergence of large variety of ST structures as will be shown in this paper.
Although in Ref. \cite{Yao2020}, the thermal noise $\xi_n(t)$ is introduced to examine spontaneous symmetry breaking, in the present work, it works merely as a seed for the ST structure formation and thus, it can be replaced by initial conditions for Eq. (\ref{Eq:FK_EOM}).

To take full account of the nonlinear effect, we numerically solve Eq. (\ref{Eq:FK_EOM}). Using the fourth-order Runge–Kutta method with the time step $\Delta t =t_{\mathrm{ex}} / 80$ ($t_{\rm ex}=2\pi/\omega_{\rm ex}$: the driving period), we integrate Eq. (\ref{Eq:FK_EOM}) starting from rest, with all initial phases and velocities being set to zero, under free boundary conditions. For the fixed parameters $\omega_0=1$, $\gamma=0.003$, $K=0.07$, and $T=0.001$, we perform observations for $N_t=3.2\times10^5$ steps after discarding $1.6\times10^6$ steps for relaxation. Since all the system parameters are normalized by powers of $\omega_0$ and $\omega_0=1$ is used, we do not explicitly write the normalizations of the parameters for simplicity.  

To identify the ST structures, we calculate the Fourier spectrum defined by 
\begin{equation}
	S(k, \omega)=\left|\frac{1}{N_x N_t}\int dt \sum_{n}\phi_n(t)e^{i(kn+\omega t)}\right|^2,
\end{equation}
where $k$ and $\omega$ denote wavenumber and frequency, respectively. We also take an ensemble average over 30 independent runs with different noise. Concerning the system size, $N_x=100$ is mainly used.
When the steady-state $S(k, \omega)$ exhibits a sharp peak at $k=k_0$ and $\omega=\omega_{\rm ex}/2$, the state is in parametric resonance with its spatial pattern is characterized by $k_0$. A threshold peak intensity distinguishing on-resonance from off-resonance cases is set to be $10^{-5}$, which is about 100 times larger than the off-resonance noise background. The steady-state ST structures shown in Fig. \ref{fig:ST} are obtained by averaging successive 10 snapshots to reduce noise disturbance.


Figure \ref{fig:Phase_diagram}(b) shows the $P_{\rm ac}$ and $\omega_{\rm ex}$ dependence of the ST structure whose wavenumber $k$ is represented by the color.
With increasing $P_{\rm ac}$ in the frequency range of $2 \lesssim \omega_{\rm ex}  \lesssim 2.25$, the steady state is changed from off-resonance (white region) to on-resonance (colored region), and the resonance wavenumber $k$ increases, suggesting that spatial patterns with various wavelength $\lambda=2\pi/k$ can be realized by tuning $P_{\rm ac}$ and $\omega_{\rm ex}$. We call the resonant state with a general value of $k$ the $k$ resonance, distinguishing it from the $\pi$ resonance shown in Fig. \ref{fig:Phase_diagram}(a). 

Figure \ref{fig:ST} shows the ST structures (left panels) and the corresponding Fourier spectra (right panels) obtained at $\omega_{\rm ex}=2.02$ for (a) $P_{\rm ac}=0.007$, (b) $P_{\rm ac}=0.024$, and (c) $P_{\rm ac}=0.05$, where in the left panels, the horizontal and vertical axes denote space (the site index $n$) and time $t/t_{\rm ex}$, respectively, and the color represents the normalized oscillation amplitude $\phi_n/\pi$.
We shall first discuss Fig. \ref{fig:ST}(a) obtained just above the parametric instability. As one can see from a red and blue checkerboard pattern in the left panel, $\phi_n$ changes its sign with the spatial and temporal periods being $\lambda\simeq 11$ and $t/t_{\rm ex}=2$, respectively. 
\begin{figure}[t]
  \centering
  \includegraphics[width=\columnwidth]{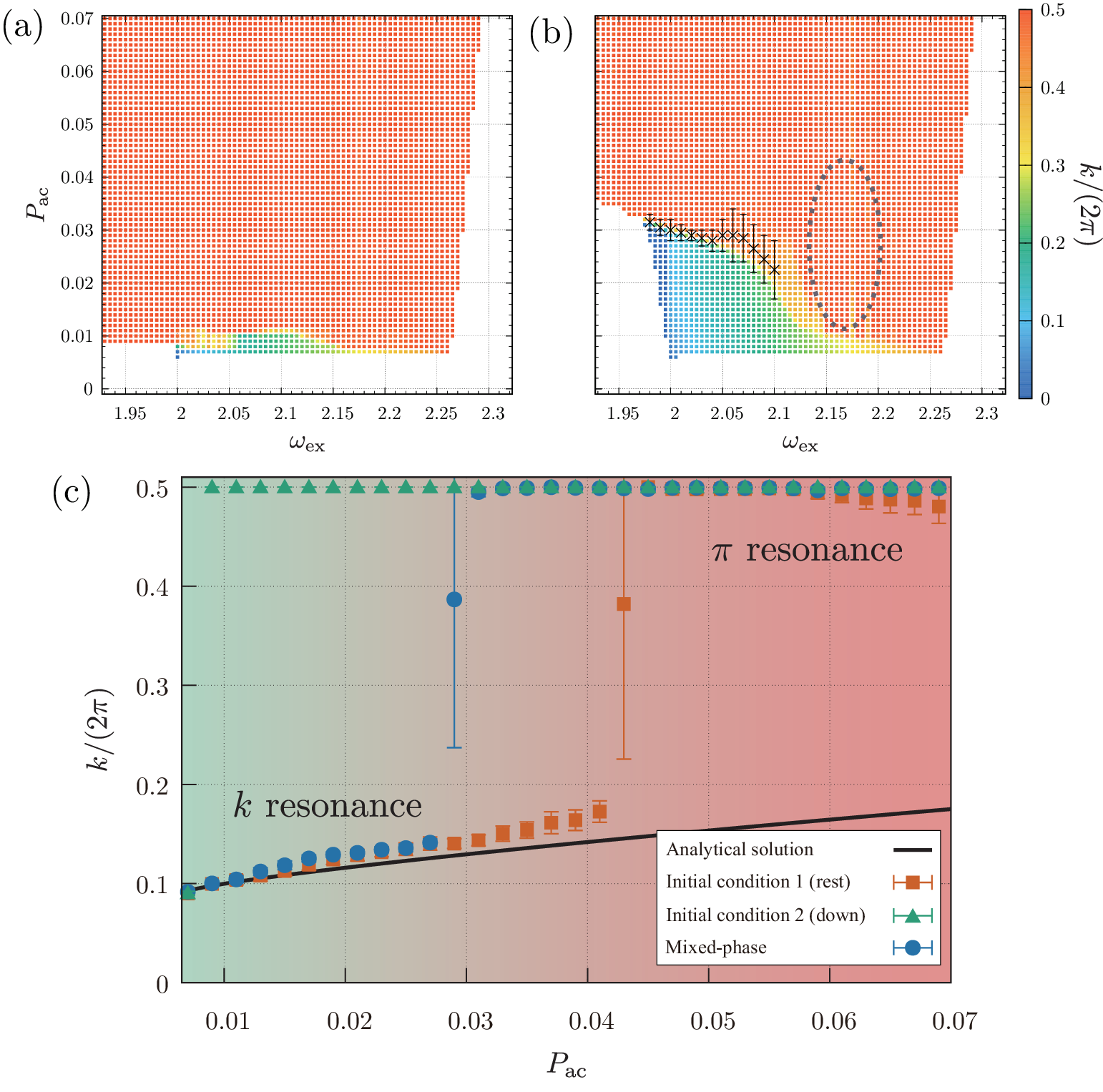}
  \caption{
  Phase diagrams obtained for the initial states of (a) the $\pi$ configuration and (b) the mixed-phase configuration, where color notations are the same as those in Fig. \ref{fig:Phase_diagram}(b). In (b), a dashed ellipse indicates the region where the mixed-phase method does not work.
  (c) $P_{\rm ac}$ dependence of the wavenumber $k$, where red, green, and blue data correspond to the numerical results shown in Figs. \ref{fig:Phase_diagram}(b), \ref{fig:selfconsistent}(a), and \ref{fig:selfconsistent}(b), respectively. The black solid line represents the analytical result obtained selfconsistently from Eqs. (\ref{Eq:Dispersion_R}) and (\ref{Eq:Single_solution}).
  }
  \label{fig:selfconsistent}
\end{figure}
The associated Fourier spectrum in the right panel exhibits a relatively sharp peak at $k /(2 \pi)=0.09$ and $2\omega/\omega_{\rm ex}=1$, being consistent with this ST structure.
As exemplified by Fig. \ref{fig:ST}(b), with increasing $P_{\mathrm{ac}}$, the wavelength $\lambda$ gets shorter and the amplitude $|\phi_n|$ becomes larger. Also, the real-space structure tends to be disturbed, which is reflected as slight broadening of the Fourier peak. This suggests that in this $P_{\rm ac}$ regime, fluctuation develops probably due to the nonlinear effect as $|\phi_n|$ is large.  
As shown in Fig. \ref{fig:ST}(c), for the sufficiently large value of $P_{\mathrm{ac}}=0.05$, there appears a ST structure with $\lambda=2$ [$k /(2 \pi)=0.5$] where $\phi_n$ on neighboring sites have opposite signs. This corresponds to the $\pi$ resonance \cite{Denardo1992,Denardo1995,Chen2007,Yao2020,Huang1993}, the stability region of which is indicated by red in Fig. \ref{fig:Phase_diagram}(a).
Although in the $\pi$ resonance, a domain wall often appears, we pick up the pure state without it in Fig. \ref{fig:ST} (c) for ease of understanding (for details, see Supplemental Material \cite{supple}).
A noteworthy aspect of the $\pi$ resonance is that the Fourier peak is very sharp and located outside the dispersion [see the right panel of Fig. \ref{fig:ST}(c)], which is in contrast to the $k$-resonance whose Fourier peak lies within the dispersion. 

Such a difference in the Fourier spectrum between the $k$ and $\pi$ resonances can be understood by considering fluctuation $\delta \phi_n$ around the parametrically resonant state $\phi_n^{\ast}$ which as shown in the left panels in Fig. \ref{fig:ST}, takes the form of a standing wave
\begin{equation}
\label{Eq:Ground_State}
\phi_n^{\ast}=A \cos \left(\frac{\omega_{\mathrm{ex}}}{2} t+\theta\right) \cos\left( kn \right)
\end{equation}
with $A$ corresponding to the maximum amplitude of $\phi_n$ in Fig. \ref{fig:ST}. Since in the right panels in Fig. \ref{fig:ST}, the background dispersion associated with the fluctuation $\delta \phi_n$ follows a similar functional form irrespective of the $k$ value, for simplicity, we shall ignore the $k$ dependence of $\phi_n^{\ast}$ in discussing $\delta \phi_n$. Then, 
by substituting $\phi_n=\phi_n^{\ast}+\delta \phi_n$ into the Eq. (\ref{Eq:FK_EOM}) and linearizing the equation with respect to $\delta \phi_n$, we obtain the dispersion relation for collective modes of the fluctuation as
\begin{equation}\label{Eq:Dispersion_R}
\begin{aligned}
&\omega_N(q)\simeq \sqrt{2 K\left(1-\cos q\right)-\frac{\gamma^2}{4}+\omega_0^2 J_0(A)}
\end{aligned}
\end{equation}
where $\delta \phi_n$ is assumed to take the following form (for details, see \cite{supple}):
\begin{equation}
	\delta \phi_n=e^{-\gamma t/2}\sum_q \Phi_q e^{i(qn+\omega_N t)}.
\end{equation}
One can see from Eq. (\ref{Eq:Dispersion_R}) that the dispersion $\omega_N(q)$ is shifted by $A$ through the Bessel function $J_0(A)$ originating from the nonlinear term. Since in Figs. \ref{fig:ST}(a)-(c), the maximum values of  
$|\phi_n| / \pi $ are $0.06$, $0.14$, and $0.5$, the associated $J_0$ values are $0.99$, $0.95$, and $0.47$, respectively. In the first two cases where $0.98 \lesssim \omega_N \lesssim 1.12$, the dispersion $\omega_N$ intersects $\omega_N=\omega_{\rm ex}/2= 1.01$, while not in the third case where $0.69 \lesssim \omega_N \lesssim 0.87$, being consistent with the numerical results in the right panels in Fig. \ref{fig:ST}. This suggests that the nonlinear effect is essential for the collective excitation. Since in the $k$ resonance, the resonant Fourier peak develops within the dispersion $\omega_N$, it is inferred that a collective mode $\omega_N$ is excited by the resonant frequency of $\omega_{\rm ex}/2$ and the parametrically amplified $A$ further shifts $\omega_N$, eventually settling down to a mode with the wavenumber $k$ determined self-consistently. In the $\pi$ resonance, such a scenario does not work, as the resonant peak is located outside the dispersion. 

A possible difference lying between the $k$ and $\pi$ resonances is also indicated from a discontinuous change in $k$ in Fig. \ref{fig:Phase_diagram}(b). This $\omega_{\rm ex}$-$P_{\rm ac}$ phase diagram is obtained by the simulation starting from rest as the initial condition. To examine the nature of the transition, we perform another simulation starting from the $\pi$ structure as the initial condition. The resultant phase diagram is shown in Fig. \ref{fig:selfconsistent}(a). One can see that the $\pi$ resonance is realized as the steady sate in the wide range of the parameter space except near the onset of the resonance. To determine which of the two states, the $k$ or $\pi$ resonance, is more likely to appear, we further perform the mixed-phase simulation \cite{Creutz1979, KA2021} in which as the initial condition, one half of the system is occupied by the $k$ structure and the other half the $\pi$ structure. Figure \ref{fig:selfconsistent}(b) shows the result obtained in the mixed-phase method. Even in the presence of the competition with the $\pi$ resonance, the $k$ resonance emerges for moderate $P_{\rm ac}$, although for larger values of $\omega_{\rm ex}$ [the region indicated by the dashed ellipse in Fig. \ref{fig:selfconsistent}(b)], this method does not work with the initial state remaining almost unchanged and thus, the relative stability cannot be determined. The $P_{\rm ac}$ dependences of $k$ obtained by the above three types of simulations are summarized in Fig. \ref{fig:selfconsistent}(c). The hysteretic behavior points to the discontinuous transition between the $k$ and $\pi$ resonant states. 

These observations naturally raise the question of whether there is a distinction in the mechanisms for the $k$ and $\pi$ resonances. To answer this question, we first consider how the wavenumber $k$ is determined. In Fig. \ref{fig:selfconsistent}(b), the $k$ value just at the onset of the resonance can easily be understood from the conventional parametric instability \cite{Chen2007,supple}. Hereafter, we will focus on inside the $k$ resonance regime where the amplitude of the phase $A \sim |\phi_n|$ is well developed. As conjectured earlier, $k$ should be selected as follows: the collective fluctuation around a $k$ resonance with amplitude $A$ is enhanced by the parametric resonance, changing into the primary resonance with a different $k$ satisfying $\omega_N(q=k)=\omega_{\rm ex}/2$. $\omega_N (q)$ in Eq. (\ref{Eq:Dispersion_R}), on the other hand, depends on the background amplitude $A$, so that this process continues until the parametric amplification of $A$ and the $A$-dependent relation $\omega_N(k)=\omega_{\rm ex}/2$ become consistent. As the information of $A$ is necessary, we shall derive a concise analytical expression of $A$ with the use of the method of multiple scales \cite{Kivshar19921001,Huang1993,Chen1994} in which the nonlinear term is approximated as $\sin \phi_n \simeq \phi_n -\frac{1}{6}\phi_n^3$ and an expansion with respect to a small parameter $\epsilon$ is performed (for details of the calculation below, see \cite{supple}).  

In the present case, we assume that all the system parameters appearing in Eqs. (\ref{Eq:FK_EOM}) and (\ref{Eq:Parametric_Driving}), $K$, $P_{\rm ac}$, and $\gamma$ are sufficiently small and can be written as $Q=\epsilon^2 Q^\prime$ ($Q=K/\omega_0^2, \, P_{\rm ac}/\omega_0^2, \, \gamma/\omega_0$) with $Q^\prime$ being $O(1)$. Under this assumption, we look for a solution of the form 
\begin{equation}
\label{Eq:phi_expand}
\phi_n(t)=\epsilon \left[a_n(\tau)e^{i\frac{\omega_{\rm ex}}{2} t}+\text{c.c.}
\right].
\end{equation}
$\phi_n$ is in the parametric resonance with frequency $\omega_{\rm ex}/2$ and its amplitude $a_n$ varies in a slow time scale $\tau=\epsilon^2 t$ separated from the fast one $t_{\rm ex}=2\pi/\omega_{\rm ex}$. For the later convenience, we introduce the Fourier transformation $a_n(\tau)=\sum_k \tilde{a}_k(\tau) e^{ikn}$. By substituting these expressions into Eq. (\ref{Eq:FK_EOM}) and picking up the nontrivial leading-order corrections of the order of $\epsilon^3$, we have
\begin{equation}\label{Eq:Single}
 \frac{d \tilde{a}_k}{d \tau} =-\frac{\gamma^\prime}{2}  \tilde{a}_k+\frac{i}{\omega_{\rm ex}}\left[\Omega^\prime_k\tilde{a}_k+\frac{P^\prime_{\rm ac}}{2} \tilde{a}^{\ast}_{-k} -\frac{\omega_0^2}{2}|(\tilde{a}_{k}|^2 + 2 |\tilde{a}^\ast_{-k}|^2)\tilde{a}_{k}\right], 
\end{equation}
where the single-mode approximation has been used and $\Omega_k^\prime$ is defined by $\Omega_k \equiv \omega_{\rm ex}(\omega_0-\frac{\omega_{\rm ex}}{2})+2K(1-\cos k) = \epsilon^2 \Omega_k^\prime$. Then, the steady-state condition $ \frac{d \tilde{a}_k}{d \tau}=0$ yields
\begin{equation}\label{Eq:Single_solution}
\begin{gathered}
|\tilde{a}_k|^2_\pm=\frac{2}{3\omega_0^2}\left[\Omega_k^\prime \pm \sqrt{\left(\frac{P^\prime_{\mathrm{ac}}}{2}\right)^2-\left(\frac{\omega_{\mathrm{ex}} \gamma^\prime}{2}\right)^2}\,\right].
\end{gathered}
\end{equation}
We note that there are two solutions, one with large amplitude $|\tilde{a}_k|_+$ and the other with small amplitude $|\tilde{a}_k|_-$. From the linear stability analysis based on Eq. (\ref{Eq:Single}), it turns out that $|\tilde{a}_k|_-$ ($|\tilde{a}_k|_+$) is unstable (stable) against small fluctuations, so that $|\tilde{a}_k|_-$ would be relevant to the $k$ resonance as it is sensitive to the fluctuation  (for details, see \cite{supple}). 

Noting the correspondence between Eqs. (\ref{Eq:Ground_State}) and (\ref{Eq:phi_expand}), we substitute $A=4 \epsilon |\tilde{a}_k|_-$ into Eq. (\ref{Eq:Dispersion_R}), and solve Eqs. (\ref{Eq:Dispersion_R}) and (\ref{Eq:Single_solution}) self-consistently to determine the wavenumber $k$ satisfying $\omega(k)=\omega_{\mathrm{ex}} / 2$. The $P_{\rm ac}$ dependence of this solution is shown by a black solid line in Fig. \ref{fig:selfconsistent}(c). One can see that in the small $P_{\rm ac}$ region, the analytically obtained wavenumber $k$ is consistent with the numerical result (colored symbols) and such a consistency is also confirmed for the amplitude $A$ (not shown), which supports the conjecture that the $k$ resonance is driven by the interplay between the parametric amplification and the collective fluctuation involving the nonlinear effect. 

In contrast to $|\tilde{a}_k|_-$, the large amplitude mode $|\tilde{a}_k|_+$ turns out not to have a self-consistent solution, so that it is irrelevant to the $k$ resonance, but instead, relevant to the $\pi$ resonance \cite{Denardo1992, Denardo_Thesis}. Actually, $|\tilde{a}_k|_+$ is basically positive defined, and thus, the occurrence of this mode is not severely restricted by the parameters values, which is consistent with the numerical result in Fig. \ref{fig:selfconsistent}(a). Also, the linear stability, i.e., the robustness against the fluctuation, can be understood from the Fourier spectrum in Fig. \ref{fig:ST}(c); due to the energy gap between the $\pi$-resonance frequency and the low-lying dispersion, the resonant state does not decay into other collective modes. All these aspects support that the $\tilde{a}_{k+}$ mode corresponds to the $\pi$ resonance. 

To conclude, we have numerically demonstrated that in the one-dimensional FK model, the parametric driving with strength $P_{\rm ac}$ and frequency $\omega_{\rm ex}$ induces spatiotemporal (ST) structures characterized by the frequency $\omega_{\rm ex}/2$ and various kinds of wavenumber $k \neq 0$. For larger $P_{\rm ac}$, the resonant state with $k=\pi$ is widely stabilized, whereas for smaller $P_{\rm ac}$, the nontrivial value of $k$ becomes possible. By further performing the extended linear stability analysis, we show that the latter $k$ resonance (the former $\pi$ resonance) corresponds to the unstable (stable) mode with a small (large) amplitude. Interestingly, the emergence of the $k$ resonance is made possible by the interplay between the parametric amplification and the collective fluctuation developing through the nonlinear effect. 

In experiments, the ST structures with $k=\pi$ and $\pi/2$ have been observed in classical pendulum systems \cite{Denardo1992, Chen2007}. Systematic research in this platform would lead to finding various types of the ST structures. Also, in the condensed matter physics, a charge density wave (CDW) state in the presence of a surface acoustic wave could be described by a similar model \cite{CDW_Fractal,CDW_SAW_Mixing,Nikitin2023}, suggesting that the CDW can be a platform for the ST physics, although the modeling should carefully be assessed \cite{Fujiwara2025}. In the CDW sliding dynamics, it is known that AC electric fields with two different frequencies lead to multi-channel interference phenomena \cite{Nikonov2021,CDW_Eac_Mixing}. 
By combining the frequency mixing with the parametric driving \cite{Porter2004}, we may have a chance to generate exotic ST patterns on various physical platforms. 
This work will advance the fundamental understanding of the nonlinear and nonequilibrium many-body effects appearing across various fields in physics, opening new possibilities for engineering via parametric resonance.       

This work was supported by Grant-in-Aid for JSPS Fellows No. JP24KJ1638 and JSPS KAKENHI Grants No. JP24K00572 and No. JP23H00257.
One of the authors (Y. F.) was supported by Program for Leading Graduate
Schools: ``Interactive Materials Science Cadet Program''.

\clearpage
\onecolumngrid
\begin{center}
  \textbf{\large Supplemental Material for\\[3pt]
  ``Spatiotemporal Structures of Parametrically Driven Nonlinear Lattices''}
\end{center}

\setcounter{section}{0}
\renewcommand{\thesection}{S\arabic{section}}
\setcounter{equation}{0}
\renewcommand{\theequation}{S\arabic{equation}}
\setcounter{figure}{0}
\renewcommand{\thefigure}{S\arabic{figure}}
\renewcommand{\thetable}{S\arabic{table}}
\makeatletter
\renewcommand{\@biblabel}[1]{[S#1]} 
\renewcommand{\@cite}[2]{[S#1\if@tempswa#2\fi]}\makeatother

\section{Domain wall and its dynamics}
\begin{figure}[b]
  \centering
\includegraphics[width=\columnwidth]{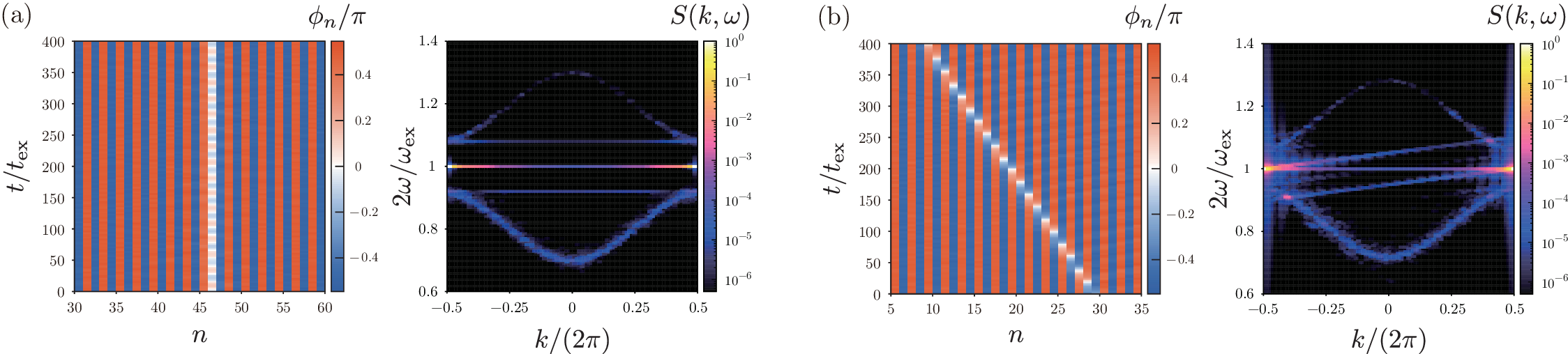}
  \caption{
  ST dynamics of a domain wall (DW) (left panels) and the corresponding Fourier spectra (right panels) for $\omega_{\mathrm{ex}}=2.02$ : (a) $P_{\mathrm{ac}}=0.068$ and (b) $P_{\mathrm{ac}}=0.04$. The notations are the same as those in Fig. \ref{fig:ST} in the main text. In the left panels, a white region corresponds to the DW.
  }
  \label{fig:DW}
\end{figure}

In the simulations, we often encounter a domain wall (DW), a phase-mismatched defect between two degenerate states, in the $\pi$-resonant state \cite{SYao2020,SDenardo1995,SDenardo1992}. Since the DW affects the Fourier spectrum, we will briefly discuss the DW dynamics.
Figure \ref{fig:DW} shows the Spatiotemporal (ST) dynamics of the DW (left panels) and its Fourier spectrum (right panels) for (a) $P_{\rm ac}=0.068$ and (b) $0.04$ at $\omega_{\rm ex}=2.02$.
For larger $P_{\rm ac}$, the DW almost stays still, which is reflected as a flat band in the Fourier spectrum [see Fig. \ref{fig:DW}(a)], whereas for smaller $P_{\rm ac}$, the DW exhibits ballistic motion, leading to an additional linear dispersion [see Fig. \ref{fig:DW}(b)]. This indicates that the dynamics of the DW can be inferred from the Fourier spectrum.

In Fig. \ref{fig:ST}(c) in the main text, we pick up a pure state without the DW for ease of understanding the typical ST structure in the $\pi$-resonant state.

\section{Dispersion relation of fluctuations in the parametrically resonant state}
We derive the dispersion relation of the fluctuation $\delta\phi_n$ around the resonant state $\phi_n^\ast$:
By substituting $\phi_n=\phi^{\ast}_n+\delta\phi_n$ into Eq. (\ref{Eq:FK_EOM}) in the main text and linearizing the equation of motion with respect to $\delta\phi_n$, we have
\begin{equation}
\ddot{\delta \phi}_n+\gamma \dot{\delta \phi}_n+\left\{\omega_0^2+P_{\mathrm{ac}} \cos \left(\omega_{\mathrm{ex}} t\right)\right\} \cos \left(\phi_n^{\ast}\right) \delta \phi_n-K\left(\delta \phi_{n+1}-2 \delta \phi_n+\delta \phi_{n-1}\right)=0.
\end{equation}
Since the resonant state exhibits a standing wave, $\phi_n^{\ast}$ takes the form of
\begin{equation}
\phi_n^{\ast}=A \cos \left(\frac{\omega_{\mathrm{ex}}}{2} t+\theta\right) \cos\left( kn\right).
\end{equation}
Since the dispersion appearing in the Fourier spectra in Fig. \ref{fig:ST} in the main text takes a similar functional form irrespective of the $k$ value, namely, the dispersion itself does not have characteristic wavenumbers, we ignore the $k$ dependence on the fluctuation $\delta \phi_n$, dropping the $\cos(kn)$ term in the above expression of $\phi_n^\ast$.    
As discussed in the main text, this approximation provides good agreement with the numerical results.
By using the formula for the Bessel function of the first kind $J_p(x)$ with integer order $p$,
\begin{equation}
\cos \left\{A \cos \left(\frac{\omega_{\mathrm{ex}}}{2} t+\theta\right)\right\}=\sum_{m=-\infty}^{\infty} J_{m}\left(A\right) \cos \left\{m\left( \frac{\omega_{\mathrm{ex}}}{2} t+\theta+\frac{\pi}{2}\right)\right\},
\end{equation}
and applying the Fourier transform $\delta\phi_n=e^{-\gamma t / 2} \sum_q \widetilde{\delta \phi}_q e^{iqn}$,
we obtain
\begin{equation}
\begin{aligned}
\sum_q&\bigg[\ddot{\widetilde{\delta \phi}}_q+\left\{2 K(1-\cos q)-\frac{\gamma^2}{4}\right\} \widetilde{\delta \phi}_q+\left\{\omega_0^2+P_{\mathrm{ac}} \cos \left(\omega_{\mathrm{ex}} t\right)\right\} \sum_{m}J_{m}\left(A\right) \cos \left\{m\left( \frac{\omega_{\mathrm{ex}}}{2} t+\theta+\frac{\pi}{2}\right)\right\}\widetilde{\delta \phi}_q
\bigg]e^{iqn}=0.
\end{aligned}
\end{equation}
Multiplying by $e^{-iq^{\ast}n}$ and summing over $n$, we obtain
\begin{equation}\label{Eq:q-space}
\begin{aligned}
&\sum_q\bigg[ \ddot{\widetilde{\delta \phi}}_q\delta_{q,q^{\ast}}+
\left\{2 K(1-\cos q)-\frac{\gamma^2}{4}\right\}\widetilde{\delta \phi}_q\delta_{q,q^{\ast}}+
\sum_{m} J_{m}\left(A\right) 
\bigg\{\frac{\omega_0^2}{2}\left(
e^{i m\left(\frac{\omega_{\mathrm{ex}}}{2} t+\theta+\frac{\pi}{2}\right)}+e^{-i m\left(\frac{\omega_{\mathrm{ex}}}{2} t+\theta+\frac{\pi}{2}\right)}\right)\\
& \quad+\frac{P_{\mathrm{ac}}}{4} \left(e^{i\left\{\frac{m\pm2}{2}\omega_{\mathrm{ex}} t+m\left(\theta+\frac{\pi}{2}\right)\right\}} +e^{-i\left\{\frac{m\mp2}{2}\omega_{\mathrm{ex}} t+m\left(\theta+\frac{\pi}{2}\right)\right\}} \right)\bigg\}\widetilde{\delta \phi}_q\delta_{q,q^*}
\bigg]=0 .
\end{aligned}
\end{equation}
Since the parametric driving yields a periodicity in time, we use the Floquet ansatz,
\begin{equation}\label{Eq:Floquet}
\widetilde{\delta \phi}_q(t)=e^{i \omega t} \sum_l \Phi_{q, l} e^{i l \omega_{\mathrm{ex}} t},
\end{equation}
where $\omega$ is response frequency distinct from the driving frequency $\omega_{\rm ex}$.
Although in general, higher harmonics of $\omega_{\rm ex}$ come into play through mode couplings, here, we consider the first zone with $l=0$ for simplicity. 
Substituting Eq. (\ref{Eq:Floquet}) into Eq. (\ref{Eq:q-space}), and integrating over time, we obtain
\begin{equation}
\begin{aligned}
& \sum_{q}\left[\left\{-\omega^2+2 K\left(1-\cos q\right)-\frac{\gamma^2}{4}\right\} \delta_{q,q^{\ast}} \delta\left(\omega-\omega^*\right)\right.
 \\
& +\sum_{m} J_m\left(A\right)\bigg\{\frac{\omega_0^2}{2}\left[e^{i m\left(\theta+\frac{\pi}{2}\right)} 
\ \delta_{q,q^{\ast}} \delta\left(\omega+m\frac{\omega_{\mathrm{ex}}}{2}-\omega^*\right)+e^{-im\left(\theta+\frac{\pi}{2}\right)} \delta_{q,q^{\ast}}  \delta\left(\omega-m\frac{\omega_{\mathrm{ex}}}{2}-\omega^*\right)\right] \\
& \left.+\frac{P_{\mathrm{ac}}}{4} \left[e^{i m\left(\theta+\frac{\pi}{2}\right)} \delta_{q,q^{\ast}} \delta\left(\omega+\frac{m \pm 2}{2} \omega_{\mathrm{ex}}-\omega^*\right)+e^{-i m\left(\theta+\frac{\pi}{2}\right)} \delta_{q,q^{\ast}}  \delta\left(\omega-\frac{m \mp 2}{2} \omega_{\mathrm{ex}}-\omega^*\right)\right]\bigg\}\right]\Phi_{q,0}=0 .
\end{aligned}
\end{equation}
By retaining only the fundamental mode, we obtain 
\begin{equation}
\left[-\left(\omega^*\right)^2+2 K\left(1-\cos q^*\right)-\frac{\gamma^2}{4}+\omega_0^2 J_0\left(A\right)-P_{\mathrm{ac}}J_2\left(A\right)\cos\left(2\theta\right)\right] \Phi_{q^\ast,0}=0.
\end{equation}
The nonlinear effect of the background resonant state with amplitude $A$ is reflected in $J_0(A)$ and $J_2(A)$. 
For the parameter set used in the numerical simulations, the $J_2(A)$ term is negligibly small compared with the $J_0(A)$ term.
Then, we finally obtain
\begin{equation}
\omega_N \simeq \pm \sqrt{2 K\left(1-\cos q\right)-\frac{\gamma^2}{4}+\omega_0^2 J_0(A)}.
\end{equation}

\section{Parametric instability and associated spatiotemporal structure}
\begin{figure}[t]
  \centering
  \includegraphics[width=\columnwidth]{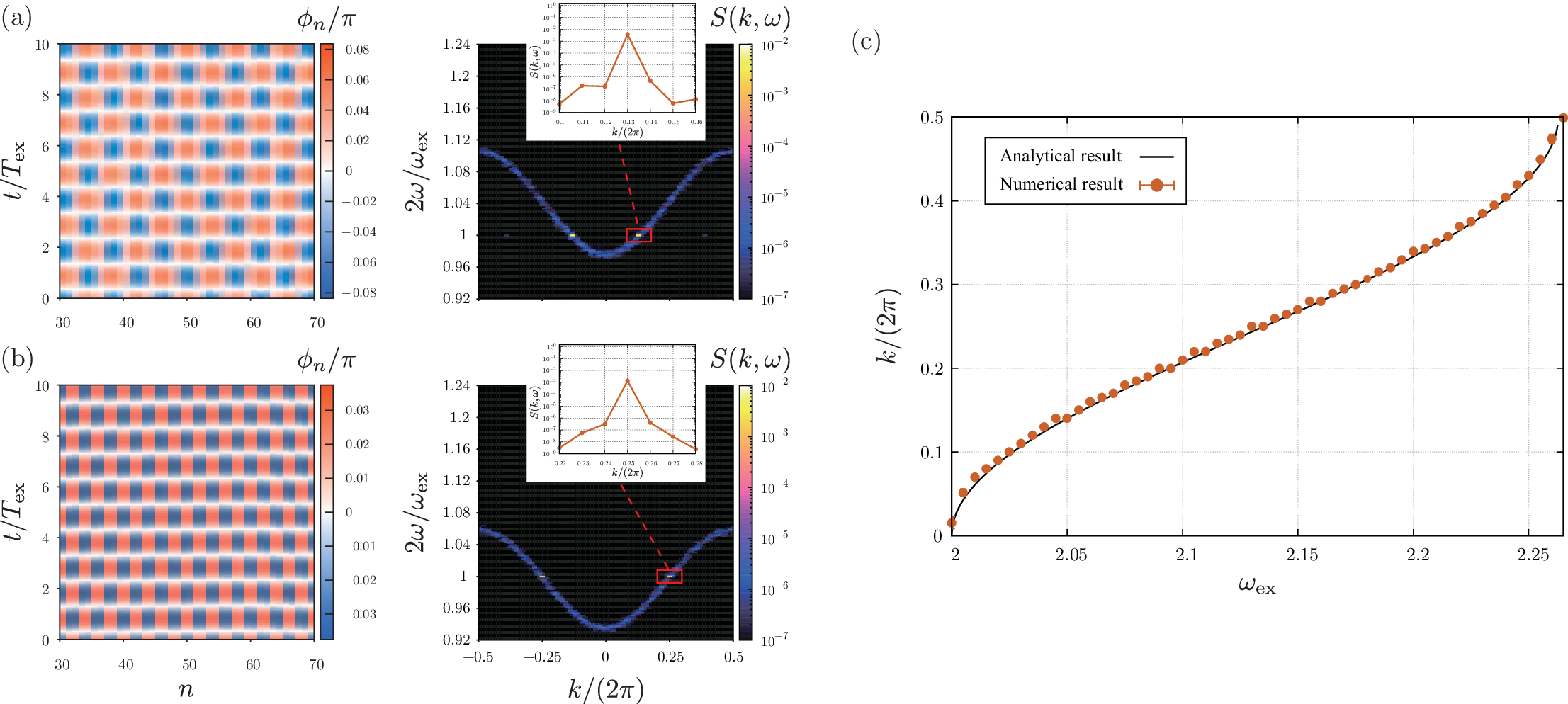}
  \caption{
  ST structures (left) and Fourier spectra (right) for (a) $\omega_{\rm ex}=2.04$ and $P_{\rm ac}=0.008$ and (b) $\omega_{\rm ex}=2.135$ and $P_{\rm ac}=0.007$ where the notations are the same manner as those in Fig. \ref{fig:ST} in the main text.
  (c) The onset wavenumber $k /(2 \pi)$ as a function of the driving frequency $\omega_{\rm ex}$ at $P_{\mathrm{ac}}=0.007$ and the other parameters are the same as those in Fig. 1(b). 
  Only at $\omega_{\rm ex}=2.0$ and 2.265 [both ends of (c)], we use $P_{\mathrm{ac}}=0.006$ and $0.01$, respectively, as the threshold values of $P_{\mathrm{ac}}$ slightly deviates from $0.007$ [see Fig. \ref{fig:Phase_diagram}(b) in the main text]. Red symbols and a black curve denote numerical and analytical results, respectively. 
  }
  \label{fig:ST_S1}
\end{figure}

In Fig. \ref{fig:ST} in the main text, we show how the wavenumber $k$ of the ST structure changes as a function of $P_{\rm ac}$. In this section, we focus on the change in $k$ just above the onset of the parametric resonance near $P_{\rm ac}=0.07$ in Fig. \ref{fig:Phase_diagram}(b) in the main text.  
Figure \ref{fig:ST_S1} shows the ST structures (left) and the corresponding Fourier spectra (right) for (a) $\omega_{\rm ex}=2.04$ and $P_{\rm ac}=0.008$ and (b) $\omega_{\rm ex}=2.135$ and $P_{\rm ac}=0.007$, both of which lie just above the parametric instability line [the border between the colored and white region in Fig. \ref{fig:Phase_diagram}(b) in the main text]. 
The ST structure in Fig. \ref{fig:ST_S1}(a) corresponds to $k/(2\pi)\simeq 0.13$ which is the same as that of Fig. \ref{fig:ST}(b) in the main text. 
Note that compared with Fig. \ref{fig:ST}(b) in the main text, where the Fourier peak is broad due to the fluctuation, the Fourier peak near the parametric instability is sharper [see the inset of the right panel of Fig. \ref{fig:ST_S1}(a)]. As the driving frequency is increased, the onset wavenumber increases, as exemplified by Fig. \ref{fig:ST_S1}(b) where the wavenumber takes the commensurate value of $k/(2\pi)\simeq 0.25$. This commensurate resonance has been reported elsewhere \cite{SDenardo1992,SKivshar19921001,SHuang1993}.
In the case of Fig. \ref{fig:ST_S1}(b), the Fourier peak is sharp (see the inset of the right panel) similarly to the case of Fig. \ref{fig:ST_S1}(a), suggesting that the fluctuation effect is relatively weak near the onset of the parametric instability.

Figure \ref{fig:ST_S1}(c) shows the $\omega_{\rm ex}$ dependence of the onset wavenumber, where red symbols denote numerical results. One can see that the wavenumber $k$ increases as the driving frequency $\omega_{\rm ex}$ deviates from $2\omega_0$.

To understand this $\omega_{\rm ex}$ dependence of the onset wavenumber, we shall examine the parametric instability.
Neglecting the nonlinear term in Eq. (\ref{Eq:FK_EOM}) in the main text, the linear equation for each Fourier mode $\delta\phi_k(t)$ in $(k,t)$-space can be obtained as
\begin{equation}\label{Eq:Damped-Mathieu}
\ddot{\delta \phi}_k+\gamma \dot{\delta \phi}_k+\left[\omega_0^2 + 2K(1-\cos k)+P_{\mathrm{ac}} \cos \left(\omega_{\mathrm{ex}} t\right)\right] \delta \phi_k=0.
\end{equation}
Since the Eq. (\ref{Eq:Damped-Mathieu}) has the same form as the damped Mathieu equation, the parametric instability, i.e, a threshold driving strength $P^{\rm (th)}_{\rm ac}$ above which a solution becomes unstable, can easily be obtained as \cite{SKovacic2018}
\begin{equation}\label{Eq:Boundary}
P_{\mathrm{ac}}^{\rm (th)}=2 \sqrt{\left(\omega_0^2+2 K(1-\cos k)-\frac{\omega_{\mathrm{ex}}^2}{4}\right)^2+\frac{\gamma^2 \omega_{\mathrm{ex}}^2}{4}}.
\end{equation}
The wavenumber that minimizes the right-hand side of Eq.  (\ref{Eq:Boundary}) is given by
\begin{equation}\label{Eq:onset_k}
k=\cos ^{-1}\left\{1+\frac{1}{2 K}\left(\omega_0^2-\frac{\omega_{\mathrm{ex}}^2}{4}\right)\right\},
\end{equation} 
which determines the wavenumber of the ST structure that first emerges, i,e., the onset wavenumber. 
The $\omega_{\rm ex}$ dependence of the analytical result Eq. (\ref{Eq:onset_k}) is plotted as a black solid curve in Fig. \ref{fig:ST_S1}(c). One can see that it is in good agreement with the numerical results.
Thus, the onset behavior can be understood within linear theory \cite{SChen2007}.
We note that $P^{\rm (th)}_{\rm ac}$ in Eq. (\ref{Eq:Boundary}) can be approximated as $P_{\rm ac}^{\rm (th)}=\gamma \omega_{\rm ex}$ for the optimal $k$ satisfying Eq. (\ref{Eq:onset_k}), which accounts for the difference between the many-body phase diagram in Fig. \ref{fig:Phase_diagram}(b) in the main text and the well-known Arnold-tongue structure in the single-body phase diagram.

\section{Method of multiple scales}
\subsection{Derivation of the equation of motion for the slow dynamics}
In this section, we provide the details of the method of multiple scales to derive Eq. (\ref{Eq:Single}) in the main text.
A fundamental assumption here is that all the system parameters appearing Eq. (\ref{Eq:FK_EOM}) in the main text are sufficiently small compared with $\omega_0^2$ and can be written with the use of a small parameter $\epsilon$ as 
\begin{equation}
K=\epsilon^2 K^{\prime},\quad P_{\mathrm{ac}}=\epsilon^2 P_{\mathrm{ac}}^{\prime}, \quad \gamma=\epsilon^2 \gamma^{\prime}, \quad \Delta=\epsilon^2 \Delta^{\prime}, 
\end{equation}
where $\Delta\equiv \omega_0-\frac{\omega_{\rm ex}}{2}$. 
Since the energy scales of $\omega_0^2$ and other parameters are quite different, associated time scales are different as well. Bearing this in our mind, we introduce a slow time variable $\tau=\epsilon^2 t$ and assume that the complex amplitude slowly varies in time with respect to the main oscillation with subharmonic resonance of $\omega_{\rm ex}/2$, namely, $\phi_n(t)=\epsilon\left[a_n(\tau)e^{i\frac{\omega_{\rm ex}}{2}t}+\text{c.c.}\right]$.
By substituting these expressions into Eq. (\ref{Eq:FK_EOM}) in the main text, one obtains the equation for a harmonic oscillator at the lowest order in $\epsilon$. Since there is no $\epsilon^2$ terms, the nontrivial leading-order contribution arises at the order in $\epsilon^3$.
 Retaining the terms of the order of $\epsilon^3$, we obtain a discrete nonlinear Schrödinger-type equation,
\begin{equation}
\frac{d a_n}{d \tau}=-\frac{\gamma^{\prime}}{2} a_n+\frac{i}{\omega_{\mathrm{ex}}}\left[\omega_{\mathrm{ex}} \Delta^{\prime} a_n-K^{\prime}\left(a_{n+1}+a_{n-1}-2 a_n\right)+\frac{P_{\mathrm{ac}}^{\prime}}{2} a_n^*-\frac{\omega_0^2}{2}\left|a_n\right|^2 a_n\right],
\end{equation}
and its complex conjugate.
With the use of the Fourier transformation $a_n(\tau)=\sum_k \tilde{a}_k(\tau) e^{i k n}$, we obtain the following expression:
\begin{equation}
\label{Eq:Mode_Coupling}
\frac{d \tilde{a}_k}{d \tau}=-\frac{\gamma^{\prime}}{2} \tilde{a}_k+\frac{i}{\omega_{\mathrm{ex}}}\left[\omega_{\mathrm{ex}} \Delta^{\prime} \tilde{a}_k+2 K^{\prime}(1-\cos k) \tilde{a}_k+\frac{P_{\mathrm{ac}}^{\prime}}{2} \tilde{a}_{-k}^*-\frac{\omega_0^2}{2} \sum_{k_1, k_2} \tilde{a}_{k_1}^{}\tilde{a}_{k_2}^* \tilde{a}_{k-k_1+k_2}^{}\right].
\end{equation}
To simplify the situation, we introduce a single-mode approximation where we focus only on the most unstable mode, i.e.,
\begin{equation}
	\sum_{k_1,k_2=\pm k}\tilde{a}_{k_1}^{} \tilde{a}_{k_2}^* \tilde{a}_{k-k_1+k_2}^{}\simeq \sum_k\left(\left|\tilde{a}_k\right|^2+2\left|\tilde{a}_{-k}\right|^2\right) \tilde{a}_k,
\end{equation}
neglecting the higher order term such as $3k$.
As the $k=\pi/2$ case is special in that $-k=-\pi / 2 \equiv 3 \pi / 2=3 k \, (\bmod 2 \pi)$, the corresponding coefficient is modified. The same is true for $k=0$ and $k=\pi$.

Now, we consider steady-state solutions of the form $\tilde{a}_{k}=A_{k}e^{i\theta_{k}}$.
Without loss of generality, $A_k$ and $\theta_k$ can be taken to satisfy $A_k=A_{-k}$ and $\theta_k=\theta_{-k}$.
As a result, we obtain the slow-flow equations for the amplitude $A_k$ and the phase $\theta_k$, given by
\begin{equation}
\label{Eq:Slow_Flow}
\left\{
\begin{aligned}
\frac{d A_k}{d \tau} & = -\frac{1}{2}\left(\gamma^{\prime}-\frac{P_{\mathrm{ac}}^{\prime}}{\omega_{\mathrm{ex}}} \sin 2 \theta_k\right) A_k \\
\frac{d \theta_k}{d \tau} & = \frac{1}{\omega_{\mathrm{ex}}}\left[\omega_{\mathrm{ex}} \Delta^{\prime}+2 K^{\prime}(1-\cos k)+\frac{P_{\mathrm{ac}}^{\prime}}{2} \cos 2 \theta_k-\frac{3 \omega_0^2}{2} A_k^2\right]
\end{aligned}
\right.
\end{equation}
The two steady-state solutions of Eq. (\ref{Eq:Slow_Flow}) correspond to Eq. (\ref{Eq:Single_solution}) in the main text.
For $k=\pi/2$ and $0, \, \pi$, the coefficient of $\omega_0^2A_k^2$ in Eq. (\ref{Eq:Slow_Flow}) is replaced with $2$ and $1/2$.

\subsection{Linear stability analysis}
As we will see below, a linear stability analysis shows that small-amplitude solutions are unstable, while large-amplitude solutions are stable.
To examine the stability, 
we introduce perturbations as $A = A_{\pm} + \delta A$ and $\theta=\theta_{\pm} + \delta \theta$ where $A_{\pm}$ and $\theta_{\pm}$ are the steady state-solutions obtained from Eq. (\ref{Eq:Slow_Flow}).
Namely, $\tilde{a}_{\pm}=A_\pm e^{i\theta_\pm}$ with
\begin{equation}\label{eq:solution}
\begin{gathered}
\cos 2 \theta_{\pm}
=\pm\sqrt{1-\left(\frac{\gamma^{\prime} \omega_{\mathrm{ex}}}{P_{\mathrm{ac}}^{\prime}}\right)^2}, 
\\
\left(A_{\pm}\right)^2=\frac{2}{3 \omega_0^2}\left[\omega_{\mathrm{ex}} \Delta^{\prime}+2 K^{\prime}(1-\cos k) \pm \frac{P_{\mathrm{ac}}^{\prime}}{2} \sqrt{1-\left(\frac{\gamma^{\prime}\omega_{\mathrm{ex}} }{P_{\mathrm{ac}}}\right)^2}\right].
\end{gathered}
\end{equation}
Here, the subscript $k$ is omitted for simplicity. Note that the small-amplitude solution $A_-$ and the large-amplitude solution $A_+$ correspond to $\cos2\theta_- <0$ and $\cos2\theta_+ >0$, respectively. Thus, the sign of $\cos2\theta_\pm$ is related to the size of the amplitude $A_\pm$.
Substituting $A = A_{\pm} + \delta A$ and $\theta=\theta_{\pm} + \delta \theta$ with Eq. (\ref{eq:solution}) back into Eq. (\ref{Eq:Slow_Flow}), we obtain 
\begin{equation}
\frac{d}{d \tau}\binom{\delta A}{\delta \theta}=\left(\begin{array}{cc}
0& \dfrac{P_{\mathrm{ac}}^{\prime}}{\omega_{\mathrm{ex}}} A_{\pm}\cos 2 \theta_{\pm} \\
-\dfrac{3 \omega_0^2}{\omega_{\mathrm{ex}}} A_{\pm} & -\gamma^{\prime}
\end{array}\right)\binom{\delta A}{\delta \theta}.
\end{equation}
Then, the eigenvalues are given by
\begin{equation}
\label{Eq:EV}
\lambda^{(\pm)}
=-\frac{\gamma^{\prime}}{2} \pm \sqrt{\left(\frac{\gamma^{\prime}}{2}\right)^2-3\left(\frac{\omega_0}{\omega_{\mathrm{ex}}} A_{\pm}\right)^2 P_{\mathrm{ac}}^{\prime} \cos 2 \theta_{\pm}}.
\end{equation}
Equation  (\ref{Eq:EV}) shows that the stability of each steady state is
determined by the sign of $P_{\mathrm{ac}}^{\prime}\cos 2\theta_{\pm}$.
The steady state is stable when $\cos 2\theta_{\pm}>0$, whereas it is unstable when $\cos 2\theta_{\pm}<0$.
Since the
large-amplitude branch $A_+$ corresponds to $\cos 2\theta_+>0$, the stability condition is always satisfied for the large-amplitude solution $\tilde{a}_{+}$ with $A_+$.
In contrast, the small-amplitude branch $A_-$ corresponds to
$\cos 2\theta_-<0$, and thus, the small-amplitude solution $\tilde{a}_{-}$ with $A_-$ is unstable.

\end{document}